\documentclass[letters,fleqn,usenatbib]{mnras}

\usepackage{newtxtext,newtxmath}

\usepackage[T1]{fontenc}

\DeclareRobustCommand{\VAN}[3]{#2}
\let\VANthebibliography\thebibliography
\def\thebibliography{\DeclareRobustCommand{\VAN}[3]{##3}\VANthebibliography}

\usepackage{graphicx}	
\usepackage{amsmath}	

\defcitealias{Mackey2019mnras}{M19}
\defcitealias{Huxor2014}{Huxor et al. 2014} 

\title[Accreted Globular Clusters]{Accreted Globular Clusters and Horizontal Branch Morphology in the Outer Halo of M31}

\author[G. McGill et al.]{Gracie McGill,$^{1}$\thanks{E-mail: g.r.mcgill@sms.ed.ac.uk (GM)}
Annette M. N. Ferguson,$^{1}$
Dougal Mackey,$^{2}$
Avon P. Huxor,$^{3}$
Geraint F. Lewis,$^{4}$
\newauthor
Nicolas F. Martin,$^{5,6}$
Alan W. McConnachie,$^{7}$
Charli M. Sakari,$^{8}$
Nial R. Tanvir,$^{9}$
Kim A. Venn,$^{10}$
\\
$^{1}$Institute for Astronomy, University of Edinburgh, Royal Observatory, Blackford Hill, Edinburgh, EH9 3HJ, UK\\
$^{2}$Independent Researcher, Charnwood, Canberra, ACT 2615, Australia\\
$^{3}$Department of Computer Science, University of Exeter, North Park Road, Exeter, EX4 4QF, UK\\
$^{4}$Sydney Institute for Astronomy, School of Physics, A28, The University of Sydney, Sydney, NSW 2006, Australia \\
$^{5}$Universit\'e de Strasbourg, CNRS, Observatoire astronomique de Strasbourg, UMR 7550, F-67000 Strasbourg, France\\
$^{6}$Max-Planck-Institut f\"{u}r Astronomie, K\"{o}nigstuhl 17, D-69117 Heidelberg, Germany\\
$^{7}$National Research Council Herzberg Astronomy and Astrophysics, 5071 West Saanich Road, Victoria, BC V9E2E7, Canada \\
$^{8}$Department of Physics \& Astronomy, San Francisco State University, San Francisco CA 94132, USA\\
$^{9}$School of Physics and Astronomy, University of Leicester, University Road, Leicester LE1 7RH, UK\\
$^{10}$Department of Physics \& Astronomy, University of Victoria, Victoria, BC V8W 3P2, Canada\\
}

\date{Accepted XXX. Received YYY; in original form ZZZ}

\pubyear{\the\year{}}

\begin{document}
\label{firstpage}
\pagerange{\pageref{firstpage}--\pageref{lastpage}}
\maketitle

\begin{abstract}
M31 hosts a rich population of outer halo ($R_{\rm{proj}} > 25$ kpc) globular clusters (GCs), many of which show strong evidence for spatial and/or kinematical associations with large-scale tidal debris features. We present deep {\it Hubble Space Telescope} photometry of 48 halo GCs, including 18 with clear ties to stellar streams and 13 with potential associations. Using the colour-magnitude diagrams (CMDs), we quantify the horizontal branch (HB) morphologies and employ new empirical relationships, calibrated on Milky Way (MW) GCs, to consistently derive metallicities and line-of-sight extinctions. We find a remarkable correlation between HB morphology and the presence of substructure: GCs with very red HBs are almost exclusively associated with substructure, while `non-substructure’ GCs have extended blue HBs. This provides the first direct evidence that red HB halo clusters originate from satellite accretion, a notion introduced nearly 50 years ago from MW studies which has remained unconfirmed until now. In addition to a more metal-rich tail, the substructure GC sample also contains a few clusters with very low metallicities and red HBs, unlike any objects known in the MW.  We suggest these are recently accreted young clusters, supporting the growing evidence that M31 has experienced a more prolonged accretion history than the MW.  
\end{abstract}

\begin{keywords}
globular clusters: general – galaxies: formation – galaxies: haloes – galaxies: individual (M31) – Local Group.
\end{keywords}



\section{Introduction}\label{intro}
Cosmological structure formation occurs hierarchically, wherein dark matter halos and the galaxies that form within them grow via merger and accretion events in which they `cannibalise' dwarf satellites. Such mergers are expected to produce an abundance of phase-space substructure in galactic stellar halos \citep[e.g.,][]{Bullock2005,Panith2021}.
A significant fraction of globular clusters (GCs) found in stellar halos are also likely to have originated in dwarf satellite galaxies that have since been accreted, a notion first proposed by \cite{SearleandZinn} in their seminal study of the GC population of the Milky Way (MW) halo. As a result, GCs associated with halo substructure present an excellent opportunity to study the accretion history of galaxies \citep[e.g.,][]{Amorisco2019,Hughes2019}. 

The challenge with this approach is identifying which halo GCs formed in situ and which have been accreted. Aside from the Sagittarius dwarf \citep{Bellazzini2020}, there is no unambiguous evidence (yet) of a MW GC embedded in a coherent tidal stream from a disrupted dwarf galaxy, so more indirect methods are required.  Over the last decade, a rich dataset has emerged for MW GCs, including detailed ages \citep[e.g.,][]{VDB2013}, orbital information \citep[e.g.,][]{Massari2019} and chemistry \citep[e.g.,][]{Horta2020}. However, this richness means that there are many ways to define in situ and accreted GC samples, and unfortunately, different methods do not always lead to the same designations \citep[e.g.,][]{Kruijssen2019,Malhan2022}. Furthermore, even when studies agree that a particular GC has been accreted, the identity of the putative host galaxy progenitor is often debated \citep[e.g.,][]{Massari2025}.   

The situation is far more straightforward in M31, where there is much more direct evidence for GC accretion. With data from the Pan-Andromeda Archaeological Survey (PAndAS; \citealt{McConnachie2018}), which mapped over 400 square degrees of sky surrounding M31, striking spatial and kinematic correlations have been observed between metal-poor outer halo substructures and GCs \citep{Mackey2010,Huxor2014,Veljanoski14,Mackey2019mnras}. These correlations have been mapped over projected radii ranging from 25--150~kpc, but there is also tentative evidence that accretion signatures exist at smaller radii as well \citep{Lewis2023}.

A comprehensive analysis of the association between 92 outer halo GCs in M31 and underlying substructure was carried out by \citet[hereafter \citetalias{Mackey2019mnras}]{Mackey2019mnras}.
They used a combination of local density and kinematic criteria to categorise each cluster as `substructure', `non-substructure' or `ambiguous'. Specifically,  a `density percentile value', $\zeta_{MP}$, was assigned to each GC, defined as the fraction of the metal-poor field halo at the projected radius of the cluster which is denser than the region local to the cluster. 
This criterion alone allowed rejection of the possibility of no correlation between GCs and substructure at the 99.95 per cent significance value. 
Folding in other considerations based on kinematics, these authors identified 32 ($\approx35$ per cent) clusters that had a high likelihood of being associated with underlying substructure (`substructure'), 35 ($\approx40$ per cent) that showed no evidence for any association (`non-substructure') and 25 where the data were ambiguous (`ambiguous').  \citet{Mackey2019Nat} analysed the kinematics of the outer halo GCs, finding that the `substructure' GCs rotate perpendicular to the `non-substructure' GCs, and argued that this was evidence of two major accretion epochs in M31. Based on the number of GCs associated with the `substructure' rotational component, they estimated the mass of the most recently-accreted progenitor to be $2 \times 10^{11} \rm{M}_{\sun}$, hence a fairly significant accretion event.  

In this Letter, we present deep {\it Hubble Space Telescope} (HST) observations of 48 GCs in the outer halo of M31, of which 18 have clear associations with stellar streams and an additional 13 are classified as having potential associations. From the colour-magnitude diagrams (CMDs), we quantify their red giant branch (RGB) and horizontal branch (HB) morphologies and employ new empirical relationships (Mackey et. al, in prep) to derive measurements of metallicity and line-of-sight extinction. With this analysis, we aim to compare the properties of `substructure' and `non-substructure' GCs.  In Section~\ref{observations} we describe the observations and data reduction, and in Section~\ref{method} we detail the methodology. Section~\ref{results} discusses the results and observed correlations between HB morphology and the degree of association with substructure. Section~\ref{CompMW} compares the overall M31 sample to the Milky Way GC population, and Section~\ref{summary} summarises the key conclusions.

\section{Observations and Data Reduction}\label{observations}

HST observations were obtained for 48 GCs in the halo of M31, of which 46 were observed with the ACS/WFC and 2 with WFC3/UVIS, as part of the following programs: HST-GO-12515, GO-13774, GO-13775 (PI:~Mackey) and GO-10394 (PI:~Tanvir). Eleven of the GCs within this sample were analysed previously by \cite{Mackey2006,Mackey2007,Perina2012,Mackey2013b} and \citet{Tanvir2012}; however, the imaging data for these have been re-reduced using the latest calibrations and consistently photometered with the rest of the sample via the improved methodology described below. Each GC was observed via 3 dithered exposures in each of the {\it F606W} and {\it F814W} filters, with typical total integration times of 2270s and 2570s, respectively.   

\begin{figure*}
\centering
\includegraphics[width=0.7\textwidth]{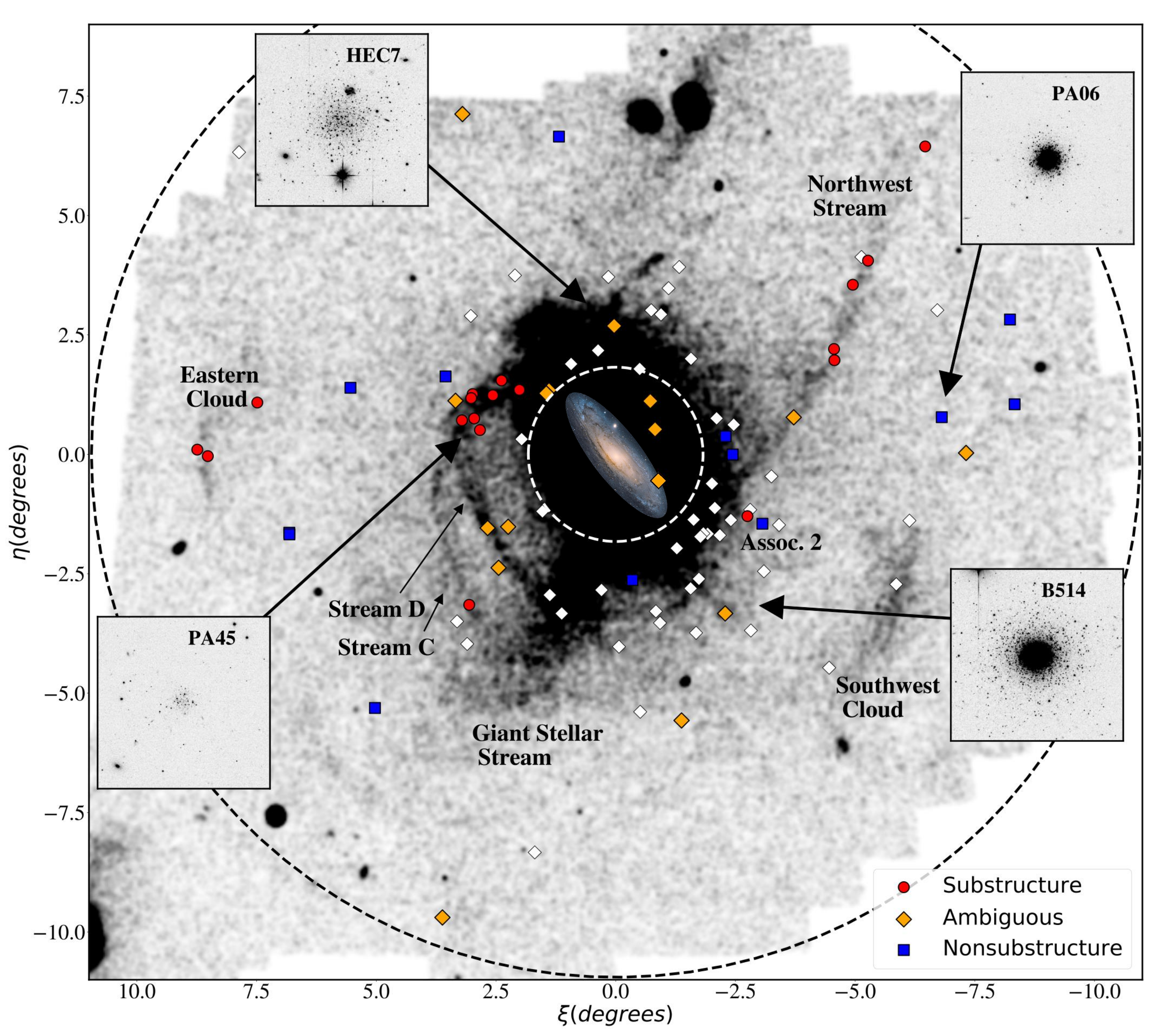}
\caption{M31 GCs on a density map of metal-poor RGB stars from the PAndAS Survey \citep{McConnachie2018}. GCs analysed herein are indicated by colour symbols according to their association with substructure \citepalias{Mackey2019mnras}. The white points indicate the known outer halo GCs of M31 that were not included in the HST sample. Inner and outer dashed circles correspond to radii of 25~kpc and 150~kpc, respectively. The inset panels show $40 \times 40$ arcsecond thumbnails for four example GCs.} 
\label{fig:GCs-map}
\end{figure*}

The positions of the observed clusters are shown in Fig.~\ref{fig:GCs-map}, coloured according to their association with the underlying substructure in the halo, as determined by \citetalias{Mackey2019mnras}. The white points indicate the outer halo GCs for which HST observations have not been obtained. Our sample contains approximately half of all known GCs in the outer halo ($R_{\rm{proj}} > 25$ kpc) of M31 and, while incomplete, spans a wide a range of magnitudes ($-9.09 \lesssim M_V \lesssim -4.06$~mag), sizes ($2.2 \lesssim r_h \lesssim 35.8$~pc) and projected radial distances (out to $R_{\rm{proj}}\sim140$~kpc) and is expected to be representative of the overall population. The inset panels in Fig.~\ref{fig:GCs-map} show thumbnails of four GCs within the sample (HEC7, PA06, PA45 and B514). These highlight the diversity of GC morphologies and sizes in M31, including several very diffuse and extended clusters \citep{Huxor2005,Huxor2011,Huxor2014, Mackey2013b}. 

For each cluster, we photometered the images using the \textsc{Dolphot} photometry software \citep{Dolphin2000,Dolphot} and followed the basic procedure described by \cite{Mackey2013b}. 
\textsc{Dolphot} performs point-spread function (PSF) fitting using model PSFs especially tailored to the relevant camera. The software provides a variety of photometric quality parameters for each detection; we selected those objects classified as stellar, with valid photometry in all input images, and global sharpness and crowding parameters falling within conservative magnitude-dependent limits, empirically determined for each target. 
Detection completeness was determined using artificial star tests.
We generated $\sim 10^5$ artificial stars per target, broadly following the distributions of high-quality stellar detections on the image and the CMD. These were added to the images and then photometered, one by one, using in-built \textsc{Dolphot} functionality. Stars were considered recovered if their measured positions were within a pixel of the input position and if they passed the quality cuts imposed on the photometry for real stars. 
We associated a completeness value with each high-quality stellar detection by considering the ratio of recovered to input artificial stars possessing radial positions and CMD locations comparable to the real star under consideration. Typical 50 per cent completeness levels were found to be $27.5$ mag in {\it F606W} and $26.8$ mag in {\it F814W}, corresponding to $\sim2-3$ magnitudes below the flat part of the HB. A full description of the observations and photometry, including a comprehensive analysis of the GC properties, will be presented in a future publication (McGill et al., in prep.). 
 
\section{Methodology}\label{method}
Cluster stars were separated from the field by imposing a radial cut. The choice of radius does not strongly affect the results presented here; it was determined on a case-by-case basis, with tighter cuts applied to clusters projecting on denser background fields (which typically sit at smaller projected radii from the centre of M31). The radii used ranged between $8-40$ arcsec.

\begin{figure}
\centering
\includegraphics[width=0.9\columnwidth]{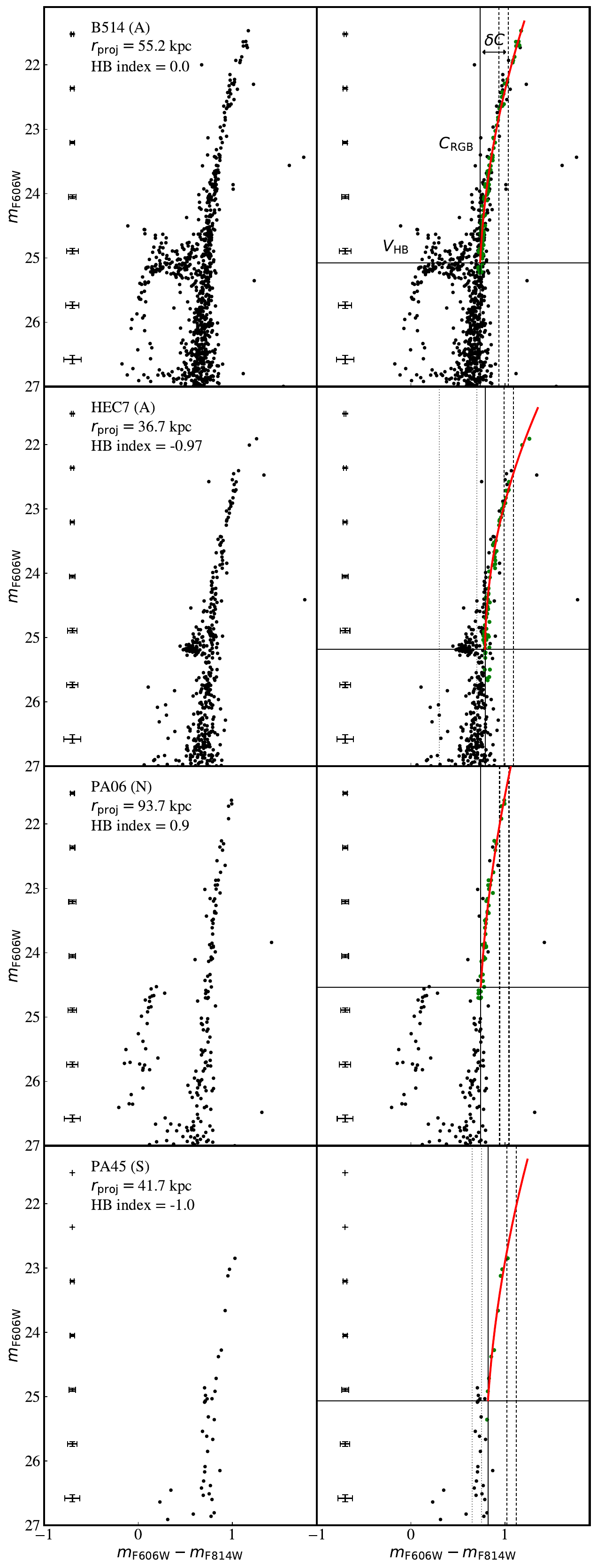}
\caption{CMDs of four example GCs. Cluster name, classification (where S = substructure, N = non-substructure and A = ambiguous), projected radius and HB index are listed in the upper left-hand corner of the left-hand panels. Key measurements are highlighted in the panels on the right: the HB magnitude, $V_{HB}$, is indicated by a solid horizontal line; the RGB colour at this magnitude, $C_{RGB}$, is likewise indicated by a solid vertical line, see labels in the top right-hand panel. The upper RGB polynomial fit is indicated by the red curve, and the two vertical dashed lines represent the $\delta C =0.2$ and $\delta C = 0.3$ intervals, respectively, with $\delta C = 0.3$ labelled explicitly.}
\label{fig:CMD-examples}
\end{figure}

Figure~\ref{fig:CMD-examples} shows the CMDs for the four example GCs highlighted previously in Fig.~\ref{fig:GCs-map}. These illustrate the variety of HB morphologies within the sample, ranging from extended blue HBs to exclusively red clump HBs. The example of PA45 (shown in the bottom panels) also highlights the sparse CMDs of the faintest and most diffuse objects. Nevertheless, it can be seen from Fig.~\ref{fig:CMD-examples} that the CMDs are of very high quality, with a $50$ per cent completeness level of $27.5$ mag in {\it F606W}, and thus the properties of their RGBs and HBs can be confidently inferred.

The panels on the right of Fig.~\ref{fig:CMD-examples} illustrate the methodology used to measure the metallicity, [Fe/H], and the line-of-sight dust extinction, $E(B-V)$, for each cluster.  This involves the measurement of several indices which are calibrated via empirical relations derived by Mackey et al. (in prep.) using observations of 47 MW GCs from the ACS Globular Cluster Treasury survey \citep[GO-10775,][]{sarajedini2007} and a smaller extension program GO-11586, \citep[GO-11586,][]{Dotter2011}. The key principle is to exploit the dependence of the upper RGB curvature on the cluster's metallicity. Automated peak-finding algorithms are employed to determine the HB magnitude, $V_{HB}$, and the RGB colour at this magnitude, $C_{RGB}$. A polynomial is then fit to the upper RGB, anchored at ($C_{RGB}$, $V_{HB}$). The curvature of the upper RGB is quantified by the difference in magnitude across a defined colour interval (either $\delta C  = 0.2$ or $\delta C = 0.3$) to the red of this anchor point, which is then used to infer [Fe/H]. Line-of-sight dust reddening, $E(B-V)$, is determined from the measurement of $C_{RGB}$. Mackey et al. (in prep.) demonstrated that this methodology reproduces the assumed literature values of the MW calibration sample very well, with a mean offset of $-0.029$~dex. 

Finally, we note that for the most metal-poor GCs, precisely determining metallicity is difficult due to the upper RGB slope becoming increasingly insensitive to metallicity in this regime. Furthermore, the spacing of the metal-poor RGB fiducial tracks becomes comparable to or smaller than the photometric errors. For nine of the M31 GCs, the metallicity determined using the above method was formally below the limit of the calibration sample ($-2.5 \la$~[Fe/H]~$\la -0.4$). Given the large uncertainties associated with these metal-poor GCs, a metallicity floor of [Fe/H] $= -2.5$ was imposed, and we present only an upper limit on their metallicities, taken to be the upper 1-$\sigma$ uncertainty associated with the measurement.

The HB morphology was also quantified via the dimensionless HB index defined by \cite{Lee1994} as $(B-R)/(B+V+R)$, where, $B$ is the number of blue HB stars, $R$ is the number of red HB stars, and $V$ is the number of stars in the instability strip. With this definition, a positive HB index represents an extended blue HB, while a negative HB index implies a red HB. The HB index of the GCs shown in Fig.~\ref{fig:CMD-examples} is labelled in the left-hand panels. To measure this index, the CMDs were de-reddened using the value of $E(B-V)$ determined previously, and a selection box was chosen to enclose the HB. The blue, red and `variable' stars were then counted (weighted by completeness), with blue stars defined here as those with $C \le 0.2$ mag, red stars as $C \ge 0.4$, and `variable' stars as those lying between these colour limits, where these limits were chosen such that the resulting HB index measurements are consistent with those presented in \cite{Mackey2005} for a MW sample (see Fig.~\ref{fig:HB-vs-met}). The uncertainties on the HB index were derived assuming Poisson statistics.   

\section{Properties of the Halo Globular Clusters}\label{results}
\begin{figure*}
\centering
\includegraphics[width=0.9\textwidth]{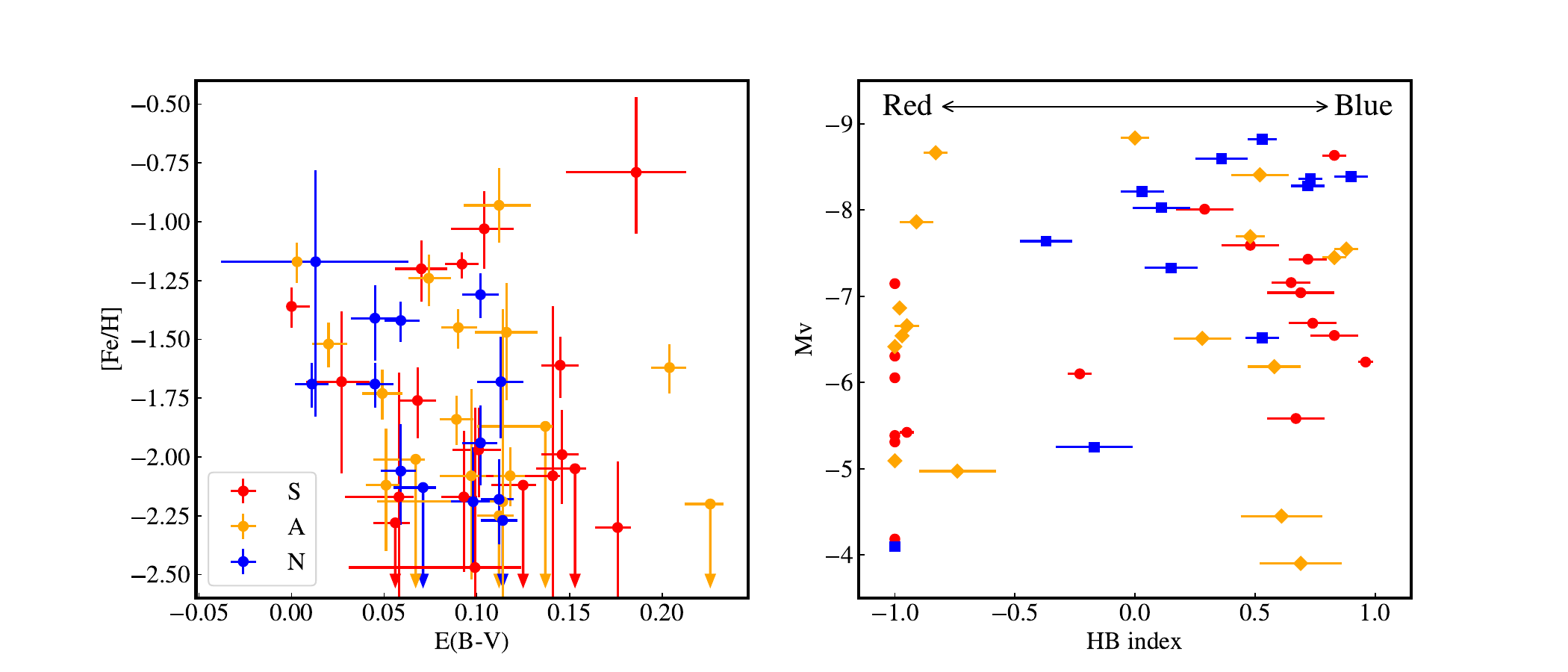}
\caption{\emph{Left:} [Fe/H] against $E(B-V)$, as derived in this study. \emph{Right:} Absolute magnitude, $M_V$, taken from \citet{Huxor2014} against HB index, measured in this study. All clusters are coloured by classification (S = substructure, N = non-substructure and A = ambiguous).}
\label{fig:all-measurements}
\end{figure*}

The left panel of Fig.~\ref{fig:all-measurements} presents our metallicity and line-of-sight extinction, $E(B-V)$, measurements. A wide range of metallicities in the sample can be seen, ranging from [Fe/H]~$\sim -0.8$ down to the imposed metallicity floor of [Fe/H]~$=-2.5$, with the metallicities of the 9 GCs hitting this floor shown as upper limits. The metallicity distributions are similar for `substructure' and `non-substructure' GCs. However, the substructure sample exhibits a more extended metal-rich tail, whereas only one non-substructure cluster (PA33) is more metal-rich than [Fe/H]~$\sim-1.3$, and this is with a large uncertainty. As expected, we see no correlation between [Fe/H] and $E(B-V)$.

\cite{Mackey2006,Mackey2007,Mackey2013b} analysed eleven HST observations in the present sample. By aligning the fiducial stellar sequences of five MW GCs presented by \cite{Brown2005} with the observed HBs and RGBs, they simultaneously estimated $E(B-V)$ and [Fe/H]. Compared with our measurements, the metallicities of \cite{Mackey2006} and \cite{Mackey2007} are systematically more metal-poor, with a mean difference of $\sim -0.4$ dex, while their $E(B-V)$ estimates are $\sim 0.05$ mag higher on average. 
Due to the nature of fitting $E(B-V)$ and [Fe/H] simultaneously, if the reddening is overestimated, then the appropriate fiducial will be misaligned on the CMD and positioned redder and fainter than the RGB stars, causing them to appear bluer by comparison and thus leading to an underestimation of the metallicity. Thus, the systematic differences are likely due to their method favouring higher $E(B-V)$ and lower [Fe/H] values, as well as the inherent uncertainty involved in interpolating between such a small set of fiducials.  Similarly, the metallicity estimated for PA48 by \cite{Mackey2013b} is also more metal-poor than measured herein; however, in this case, the values agree within their respective uncertainties. 

Detailed abundance measurements via high-resolution integrated light (IL) spectroscopy have previously been presented for several GCs in the outer halo of M31. \citet{Colucci2014} present metallicities for two GCs in our present sample, which are in good agreement with our values. 
Six GCs from our sample were also analysed by \cite{Sakari2015} and we find our metallicities are $0.37$~dex lower on average, though this effect is less prominent at the metal-rich end.
Additionally,  \cite{Sakari2022} use IL spectroscopy of the Calcium-II triplet (CaT) to estimate metallicity of thirteen of the GCs analysed herein, and we find that, on average, our values are $0.15$~dex more metal-poor.  Finally, the CaT metallicities presented by \cite{Usher2024} for sixteen GCs in our sample are in excellent agreement with our results, with a mean difference of $-0.005$~dex. We also note that four GCs hitting the imposed metallicity floor are included in the analyses of \cite{Sakari2015}, \cite{Sakari2022} and \cite{Usher2024}. Encouragingly, these GCs are all found to have spectroscopic metallicities of [Fe/H]~$\lesssim-2.0$, supporting the case for these clusters being genuinely very metal-poor. 

The right-hand panel of Fig.~\ref{fig:all-measurements} presents absolute magnitude, taken from  \cite{Huxor2014}, against the measured HB index.
A large spread in HB morphology is apparent in the outer halo, but what is immediately striking is the difference between the HB morphologies of those clusters that are and are not associated with substructure. Remarkably, all but one of the clusters with red HBs (defined here as HB index~$<-0.5$) have either strong or potential links to substructure. The only exception to this is PA38, with an HB index = $-1.0$, which is a particularly faint ($M_V = -4.50$) and extended ($r_h = 24.4$ pc) cluster \citep{Huxor2014} lying at a projected radius of 92~kpc in the outer halo. 

Additionally, in Fig.~\ref{fig:all-measurements} we see potential evidence that GCs with red HBs are fainter on average than those with blue HBs, with a mean magnitude of~$-6.0$ mag for red-HB GCs, compared to $-7.3$ mag for GCs with blue HBs. Indeed, a two-sample K-S test gives a probability of $\sim 0.1$ per cent that these samples were drawn from the same underlying distribution. However, this effect may stem from a slight bias within the `non-substructure' sample. While our `non-substructure' GC sample spans the full magnitude range of known non-substructure GCs in the outer halo, the fainter peak of the bimodal luminosity distribution \citepalias{Huxor2014,Mackey2019mnras}, located at $M_V\approx -5.4$ \citepalias{Mackey2019mnras}, is not particularly well sampled in the present study. This, in combination with the observation that non-substructure GCs exhibit blue HB morphologies, may explain why blue-HBs GCs appear brighter on average.

\section{Comparison to the Milky Way}\label{CompMW}
\begin{figure*}
\centering
\includegraphics[width=0.8\textwidth]{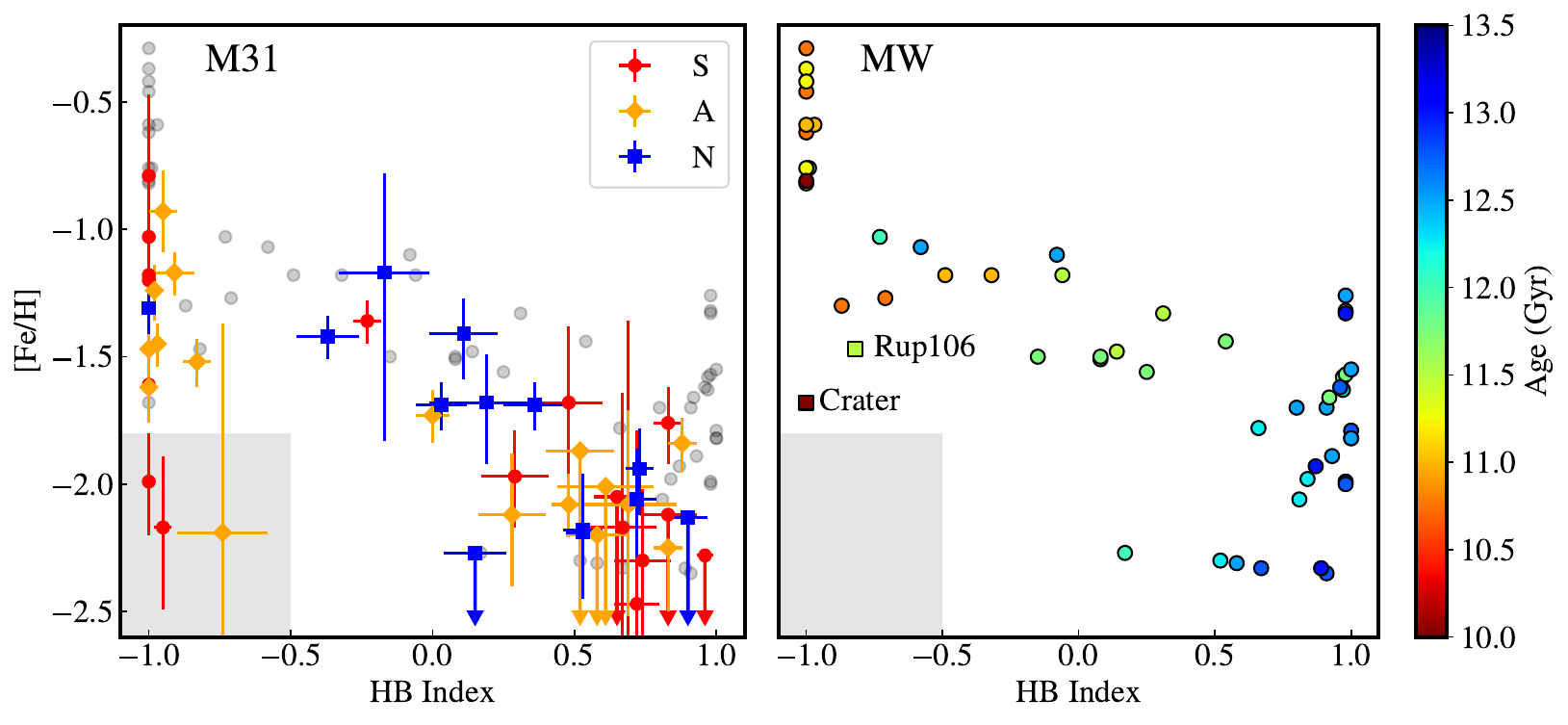}
\caption{Metallicity vs HB morphology. \emph{Left:} M31 GCs, coloured by association with substructure (S = substructure, N = non-substructure and A = ambiguous). MW GCs are plotted as light grey circles. \emph{Right:} MW GCs are presented as circles coloured by cluster age, Ruprecht 106 and Crater are shown as square symbols.}
\label{fig:HB-vs-met}
\end{figure*}

Figure~\ref{fig:HB-vs-met} illustrates the relationship between metallicity and HB morphology. The M31 GCs analysed here are shown in the left panel, while the right panel shows a sample of MW GCs, where their metallicities are taken from the spectroscopic measurements made by \cite{Carretta2009}\footnote{With the exceptions of Ruprecht~106 \citep{Villanova2013,Dotter2011} and Crater \citep{Kirby2015, Weisz2016}, highlighted as square symbols.}. These metallicity measurements were used to calibrate the empirical relationships derived by Mackey et al. (in prep), so the MW and M31 GC metallicities are on the same scale and can be directly compared. HB indices for MW GCs are taken from \cite{Mackey2005}, and, as mentioned previously, the method for quantifying HB morphology herein was defined such that these results are consistent.  

In the left panel, the difference between the `substructure' and `non-substructure' GCs is again highlighted, with `substructure' clusters extending to slightly higher metallicities and dominating at the reddest HB morphologies, compared to the near-exclusively blue `non-substructure' GCs. Overall, the M31 sample broadly follows the same trend between HB morphology and metallicity that is seen in the MW. That is, the most metal-poor GCs typically exhibit blue HBs while red HB GCs are generally more metal-rich. We note that the M31 sample appears to lack the most metal-rich tail seen in the MW population, however, this is almost certainly an artefact of our sample containing only outer halo GCs, while the MW sample contains metal-rich bulge clusters.

In their seminal study of the MW GC population, \cite{SearleandZinn} used the spread in HB morphologies at a given metallicity to argue for a prolonged period of accretion in building up the Galactic outer halo. This notion has been reinforced through many later studies \citep[e.g.,][]{Zinn93, Mackey2005}, with compelling evidence put forth for red-HB outer halo GCs having somewhat younger ages \citep[e.g.,][]{Stetson1999, Dotter2010, Dotter2011, Weisz2016}. This can be seen in the right panel of Fig.~\ref{fig:HB-vs-met}, where MW GCs are coloured according to their main sequence turn-off ages derived by \cite{VDB2013}. While this is highly suggestive of these clusters being late arrivals, the `smoking gun' of GC accretion in the MW halo has been difficult to obtain.  In M31, tantalising evidence of an association between red-HB GCs and recent accretion was previously presented by \cite{Mackey2013}, who imaged PA07 and PA08, two GCs associated with the South-West Cloud (see fig.~\ref{fig:GCs-map}),
with the Gemini Multi-Object Spectrograph and characterised their CMDs. Through comparison to MW GCs with the same (red) HB morphology and metallicity, \cite{Mackey2013} concluded that both of these `substructure' GCs were likely to be at least 2~Gyr younger than the oldest clusters observed in the MW.  
Using our much larger sample of M31 GCs, which probes a broader range of halo substructures, we provide the first incontrovertible evidence that red-HB halo GCs are accreted.  

A previous analysis of the HB morphology of M31 halo GCs was presented by \cite{Perina2012}, who found that they  had redder HBs at a given metallicity than their MW counterparts. They interpreted this as suggestive that these GCs are, on average, 1~Gyr younger than those found in the halo of the MW. While such a global difference between the two populations is not as evident in our larger study which
focuses exclusively on the outer halo, we do confirm that some M31 outer halo GCs have considerably redder HBs than their MW counterparts at the same metallicity. 

The red-HB GCs clearly originated in one or more satellite progenitors that presumably underwent more extended star formation histories before being accreted onto the M31 halo. Knowledge of the ages of these clusters would therefore be extremely valuable for constraining the timescale on which these accretion events took place. Unfortunately, precise ages are not available for any of the M31 GCs in our sample. Crude ages of 19 of these GCs were recently estimated by \cite{Usher2024} who used stellar population model fits to integrated light photometry from PAndAS, invoking a metallicity prior from the near-infrared CaT spectral feature.  While they find marginal evidence that the `substructure' and `ambiguous' GCs are slightly younger on average than those of the smooth halo (11.3~Gyr vs 12.4~Gyr), this difference lies within their significant measurement uncertainties ($\sim 2$ Gyr).  

 Figure~\ref{fig:HB-vs-met} shows that there are three M31 outer halo GCs with red HBs and very low metallicities ([Fe/H] $\le -1.8$), highlighted with a shaded region. These objects have no counterparts in the MW, with the closest MW cluster in this parameter space being Crater -- a much younger ($7.5$ Gyr) cluster at a Galactocentric radius of $\sim145$~kpc which has been hypothesised to have been accreted \citep{Weisz2016}. Notably, all three of the unusual M31 GCs have links with substructure. Two have strong associations: PA45 is associated with the overlapping region of streams C and D, and PA18 is part of association 2 (an overdensity of 11 tightly grouped GCs first noted by \citealt{Mackey2010}). PA15 has potential links to the north-west stream.
Given their low metallicities and very red HBs, it is tempting to speculate that they have been donated by one or more low-mass satellites that remained actively star-forming until rather recently. Indeed, the absence of such objects in the MW supports the growing evidence that M31 has experienced a much more extended accretion history, with a significant event occurring in the last few Gyr \citep{Bernard2015, Hammer2018, Mackey2019Nat}.  

It is worth briefly acknowledging the possibility that these metallicities may be underestimated. The empirical relationships that we have adopted to determine metallicities were calibrated using mostly old Galactic GCs with enhanced $\alpha$-abundances relative to solar ([$\alpha$/Fe]$\approx +0.4$) and so the metallicities of GCs whose properties deviate from these could be underestimated. Using theoretical isochrones from the BaSTI database \citep{Pietrinferni2021}, Mackey et al. (in prep.) estimate that for a cluster with [$\alpha$/Fe]~$\approx 0.0$, the metallicity may be underestimated by as much as $\sim0.3$~dex. However, when applied to Rup~106, which has a similarly old age to the calibration GCs but with a lower abundance of $\alpha$-elements ([$\alpha$/Fe]~$\approx 0.0$, \citealp{Villanova2013}), the metallicity determined by this methodology in agreement with the values from spectroscopy, suggesting that, in this case at least, the effect of a reduced $\alpha$-abundance alone may not be as significant as suggested by theoretical isochrones.

\section{Summary and Conclusions}\label{summary}

In this Letter, we have presented deep HST observations of 48 GCs in the halo of M31, of which 18 have strong links to tidal debris features and 13 have potential associations. From the CMDs, we quantify the HB morphology and employ a new methodology, based on empirical relationships calibrated with MW GCs, to consistently derive measurements of the metallicity and line-of-sight extinction. We find the remarkable result that GCs with red HBs (HB index $\lesssim -0.5$) are almost exclusively linked to substructure.  
While `substructure' and `ambiguous' GCs exhibit a range of HB morphologies, all `non-substructure' GCs are found to have extended blue HBs. We interpret these results as the first direct evidence that red-HB GCs in galactic halos are accreted, supporting the scenario initially suggested by \cite{SearleandZinn} from MW studies several decades ago.  Additionally, we find 3 GCs with extremely red HBs and very low metallicities, unlike any known objects in the MW, which could be examples of rather young GCs, which have been accreted recently from low mass satellites. 

The sample analysed in this study contains only half of the known GCs in the outer halo of M31 and obtaining observations of the remaining outer halo cluster population will be crucial for confirming our results. In addition,  it will be extremely important to obtain deep main sequence turnoff photometry to establish the precise ages of the accreted red-HB GCs and confirm the hypothesis that they are young. Our future work will present similar analyses of the GC populations of Local Group dwarf galaxies to understand how the accretions of such objects might contribute to halos of the MW and M31, and to search for counterparts to the very metal-poor red HB GCs identified in this work. 

\section*{Acknowledgements}

GM acknowledges funding from the Bell Burnell Graduate Scholarship Fund [grant number BB0027]. 
AMNF is supported by UK Research and Innovation (UKRI) under the UK government’s Horizon Europe funding guarantee [grant number EP/Z534353/1] and by the UK Science and Technology Facilities
Council [grant number ST/Y001281/1].

For the purpose of open access, the author has applied a Creative Commons Attribution (CC BY)
licence to any Author Accepted Manuscript version arising from this submission

\section*{Data Availability}

The data used in this analysis is available from the Mikulski Archive for Space Telescopes (MAST) with the programme IDs GO-10394, GO-12515, GO-13774 and GO-13775.



\bibliographystyle{mnras}
\bibliography{m31letter} 

\begin{thebibliography}{}
\makeatletter
\relax
\def\mn@urlcharsother{\let\do\@makeother \do\$\do\&\do\#\do\^\do\_\do\%\do\~}
\def\mn@doi{\begingroup\mn@urlcharsother \@ifnextchar [ {\mn@doi@} {\mn@doi@[]}}
\def\mn@doi@[#1]#2{\def\@tempa{#1}\ifx\@tempa\@empty \href {http://dx.doi.org/#2} {doi:#2}\else \href {http://dx.doi.org/#2} {#1}\fi \endgroup}
\def\mn@eprint#1#2{\mn@eprint@#1:#2::\@nil}
\def\mn@eprint@arXiv#1{\href {http://arxiv.org/abs/#1} {{\tt arXiv:#1}}}
\def\mn@eprint@dblp#1{\href {http://dblp.uni-trier.de/rec/bibtex/#1.xml} {dblp:#1}}
\def\mn@eprint@#1:#2:#3:#4\@nil{\def\@tempa {#1}\def\@tempb {#2}\def\@tempc {#3}\ifx \@tempc \@empty \let \@tempc \@tempb \let \@tempb \@tempa \fi \ifx \@tempb \@empty \def\@tempb {arXiv}\fi \@ifundefined {mn@eprint@\@tempb}{\@tempb:\@tempc}{\expandafter \expandafter \csname mn@eprint@\@tempb\endcsname \expandafter{\@tempc}}}

\bibitem[\protect\citeauthoryear{{Amorisco}}{{Amorisco}}{2019}]{Amorisco2019}
{Amorisco} N.~C.,  2019, \mn@doi [\mnras] {10.1093/mnras/sty2927}, \href {https://ui.adsabs.harvard.edu/abs/2019MNRAS.482.2978A} {482, 2978}

\bibitem[\protect\citeauthoryear{{Bellazzini}, {Ibata}, {Malhan}, {Martin}, {Famaey}  \& {Thomas}}{{Bellazzini} et~al.}{2020}]{Bellazzini2020}
{Bellazzini} M.,  {Ibata} R.,  {Malhan} K.,  {Martin} N.,  {Famaey} B.,   {Thomas} G.,  2020, \mn@doi [\aap] {10.1051/0004-6361/202037621}, \href {https://ui.adsabs.harvard.edu/abs/2020A&A...636A.107B} {636, A107}

\bibitem[\protect\citeauthoryear{{Bernard} et~al.,}{{Bernard} et~al.}{2015}]{Bernard2015}
{Bernard} E.~J.,  et~al., 2015, \mn@doi [\mnras] {10.1093/mnras/stu2309}, \href {https://ui.adsabs.harvard.edu/abs/2015MNRAS.446.2789B} {446, 2789}

\bibitem[\protect\citeauthoryear{{Brown} et~al.,}{{Brown} et~al.}{2005}]{Brown2005}
{Brown} T.~M.,  et~al., 2005, \mn@doi [\aj] {10.1086/444542}, \href {https://ui.adsabs.harvard.edu/abs/2005AJ....130.1693B} {130, 1693}

\bibitem[\protect\citeauthoryear{{Bullock} \& {Johnston}}{{Bullock} \& {Johnston}}{2005}]{Bullock2005}
{Bullock} J.~S.,  {Johnston} K.~V.,  2005, \mn@doi [\apj] {10.1086/497422}, \href {https://ui.adsabs.harvard.edu/abs/2005ApJ...635..931B} {635, 931}

\bibitem[\protect\citeauthoryear{{Carretta}, {Bragaglia}, {Gratton}, {D'Orazi}  \& {Lucatello}}{{Carretta} et~al.}{2009}]{Carretta2009}
{Carretta} E.,  {Bragaglia} A.,  {Gratton} R.,  {D'Orazi} V.,   {Lucatello} S.,  2009, \mn@doi [\aap] {10.1051/0004-6361/200913003}, \href {https://ui.adsabs.harvard.edu/abs/2009A&A...508..695C} {508, 695}

\bibitem[\protect\citeauthoryear{{Colucci}, {Bernstein}  \& {Cohen}}{{Colucci} et~al.}{2014}]{Colucci2014}
{Colucci} J.~E.,  {Bernstein} R.~A.,   {Cohen} J.~G.,  2014, \mn@doi [\apj] {10.1088/0004-637X/797/2/116}, \href {https://ui.adsabs.harvard.edu/abs/2014ApJ...797..116C} {797, 116}

\bibitem[\protect\citeauthoryear{{Dolphin}}{{Dolphin}}{2000}]{Dolphin2000}
{Dolphin} A.~E.,  2000, \mn@doi [\pasp] {10.1086/316630}, \href {https://ui.adsabs.harvard.edu/abs/2000PASP..112.1383D} {112, 1383}

\bibitem[\protect\citeauthoryear{{Dolphin}}{{Dolphin}}{2016}]{Dolphot}
{Dolphin} A.,  2016, {DOLPHOT: Stellar photometry}, Astrophysics Source Code Library, record ascl:1608.013

\bibitem[\protect\citeauthoryear{{Dotter} et~al.,}{{Dotter} et~al.}{2010}]{Dotter2010}
{Dotter} A.,  et~al., 2010, \mn@doi [\apj] {10.1088/0004-637X/708/1/698}, \href {https://ui.adsabs.harvard.edu/abs/2010ApJ...708..698D} {708, 698}

\bibitem[\protect\citeauthoryear{{Dotter}, {Sarajedini}  \& {Anderson}}{{Dotter} et~al.}{2011}]{Dotter2011}
{Dotter} A.,  {Sarajedini} A.,   {Anderson} J.,  2011, \mn@doi [\apj] {10.1088/0004-637X/738/1/74}, \href {https://ui.adsabs.harvard.edu/abs/2011ApJ...738...74D} {738, 74}

\bibitem[\protect\citeauthoryear{{Hammer}, {Yang}, {Wang}, {Ibata}, {Flores}  \& {Puech}}{{Hammer} et~al.}{2018}]{Hammer2018}
{Hammer} F.,  {Yang} Y.~B.,  {Wang} J.~L.,  {Ibata} R.,  {Flores} H.,   {Puech} M.,  2018, \mn@doi [\mnras] {10.1093/mnras/stx3343}, \href {https://ui.adsabs.harvard.edu/abs/2018MNRAS.475.2754H} {475, 2754}

\bibitem[\protect\citeauthoryear{{Horta} et~al.,}{{Horta} et~al.}{2020}]{Horta2020}
{Horta} D.,  et~al., 2020, \mn@doi [\mnras] {10.1093/mnras/staa478}, \href {https://ui.adsabs.harvard.edu/abs/2020MNRAS.493.3363H} {493, 3363}

\bibitem[\protect\citeauthoryear{{Hughes}, {Pfeffer}, {Martig}, {Bastian}, {Crain}, {Kruijssen}  \& {Reina-Campos}}{{Hughes} et~al.}{2019}]{Hughes2019}
{Hughes} M.~E.,  {Pfeffer} J.,  {Martig} M.,  {Bastian} N.,  {Crain} R.~A.,  {Kruijssen} J.~M.~D.,   {Reina-Campos} M.,  2019, \mn@doi [\mnras] {10.1093/mnras/sty2889}, \href {https://ui.adsabs.harvard.edu/abs/2019MNRAS.482.2795H} {482, 2795}

\bibitem[\protect\citeauthoryear{{Huxor}, {Tanvir}, {Irwin}, {Ibata}, {Collett}, {Ferguson}, {Bridges}  \& {Lewis}}{{Huxor} et~al.}{2005}]{Huxor2005}
{Huxor} A.~P.,  {Tanvir} N.~R.,  {Irwin} M.~J.,  {Ibata} R.,  {Collett} J.~L.,  {Ferguson} A.~M.~N.,  {Bridges} T.,   {Lewis} G.~F.,  2005, \mn@doi [\mnras] {10.1111/j.1365-2966.2005.09086.x}, \href {https://ui.adsabs.harvard.edu/abs/2005MNRAS.360.1007H} {360, 1007}

\bibitem[\protect\citeauthoryear{{Huxor} et~al.,}{{Huxor} et~al.}{2011}]{Huxor2011}
{Huxor} A.~P.,  et~al., 2011, \mn@doi [\mnras] {10.1111/j.1365-2966.2011.18450.x}, \href {https://ui.adsabs.harvard.edu/abs/2011MNRAS.414..770H} {414, 770}

\bibitem[\protect\citeauthoryear{{Huxor} et~al.,}{{Huxor} et~al.}{2014}]{Huxor2014}
{Huxor} A.~P.,  et~al., 2014, \mn@doi [\mnras] {10.1093/mnras/stu771}, \href {https://ui.adsabs.harvard.edu/abs/2014MNRAS.442.2165H} {442, 2165}

\bibitem[\protect\citeauthoryear{{Kirby}, {Simon}  \& {Cohen}}{{Kirby} et~al.}{2015}]{Kirby2015}
{Kirby} E.~N.,  {Simon} J.~D.,   {Cohen} J.~G.,  2015, \mn@doi [\apj] {10.1088/0004-637X/810/1/56}, \href {https://ui.adsabs.harvard.edu/abs/2015ApJ...810...56K} {810, 56}

\bibitem[\protect\citeauthoryear{{Kruijssen}, {Pfeffer}, {Reina-Campos}, {Crain}  \& {Bastian}}{{Kruijssen} et~al.}{2019}]{Kruijssen2019}
{Kruijssen} J.~M.~D.,  {Pfeffer} J.~L.,  {Reina-Campos} M.,  {Crain} R.~A.,   {Bastian} N.,  2019, \mn@doi [\mnras] {10.1093/mnras/sty1609}, \href {https://ui.adsabs.harvard.edu/abs/2019MNRAS.486.3180K} {486, 3180}

\bibitem[\protect\citeauthoryear{{Lee}, {Demarque}  \& {Zinn}}{{Lee} et~al.}{1994}]{Lee1994}
{Lee} Y.-W.,  {Demarque} P.,   {Zinn} R.,  1994, \mn@doi [\apj] {10.1086/173803}, \href {https://ui.adsabs.harvard.edu/abs/1994ApJ...423..248L} {423, 248}

\bibitem[\protect\citeauthoryear{{Lewis}, {Brewer}, {Mackey}, {Ferguson}, {Li}  \& {Adams}}{{Lewis} et~al.}{2023}]{Lewis2023}
{Lewis} G.~F.,  {Brewer} B.~J.,  {Mackey} D.,  {Ferguson} A. M.~N.,  {Li} Y.~C.,   {Adams} T.,  2023, \mn@doi [\mnras] {10.1093/mnras/stac3325}, \href {https://ui.adsabs.harvard.edu/abs/2023MNRAS.518.5778L} {518, 5778}

\bibitem[\protect\citeauthoryear{{Mackey} \& {van den Bergh}}{{Mackey} \& {van den Bergh}}{2005}]{Mackey2005}
{Mackey} A.~D.,  {van den Bergh} S.,  2005, \mn@doi [\mnras] {10.1111/j.1365-2966.2005.09080.x}, \href {https://ui.adsabs.harvard.edu/abs/2005MNRAS.360..631M} {360, 631}

\bibitem[\protect\citeauthoryear{{Mackey} et~al.,}{{Mackey} et~al.}{2006}]{Mackey2006}
{Mackey} A.~D.,  et~al., 2006, \mn@doi [\apjl] {10.1086/510670}, \href {https://ui.adsabs.harvard.edu/abs/2006ApJ...653L.105M} {653, L105}

\bibitem[\protect\citeauthoryear{{Mackey} et~al.,}{{Mackey} et~al.}{2007}]{Mackey2007}
{Mackey} A.~D.,  et~al., 2007, \mn@doi [\apjl] {10.1086/511977}, \href {https://ui.adsabs.harvard.edu/abs/2007ApJ...655L..85M} {655, L85}

\bibitem[\protect\citeauthoryear{{Mackey} et~al.,}{{Mackey} et~al.}{2010}]{Mackey2010}
{Mackey} A.~D.,  et~al., 2010, \mn@doi [\apjl] {10.1088/2041-8205/717/1/L11}, \href {https://ui.adsabs.harvard.edu/abs/2010ApJ...717L..11M} {717, L11}

\bibitem[\protect\citeauthoryear{{Mackey} et~al.,}{{Mackey} et~al.}{2013a}]{Mackey2013}
{Mackey} A.~D.,  et~al., 2013a, \mn@doi [\mnras] {10.1093/mnras/sts336}, \href {https://ui.adsabs.harvard.edu/abs/2013MNRAS.429..281M} {429, 281}

\bibitem[\protect\citeauthoryear{{Mackey} et~al.,}{{Mackey} et~al.}{2013b}]{Mackey2013b}
{Mackey} A.~D.,  et~al., 2013b, \mn@doi [\apjl] {10.1088/2041-8205/770/2/L17}, \href {https://ui.adsabs.harvard.edu/abs/2013ApJ...770L..17M} {770, L17}

\bibitem[\protect\citeauthoryear{{Mackey} et~al.,}{{Mackey} et~al.}{2019a}]{Mackey2019mnras}
{Mackey} A.~D.,  et~al., 2019a, \mn@doi [\mnras] {10.1093/mnras/stz072}, \href {https://ui.adsabs.harvard.edu/abs/2019MNRAS.484.1756M} {484, 1756}

\bibitem[\protect\citeauthoryear{{Mackey} et~al.,}{{Mackey} et~al.}{2019b}]{Mackey2019Nat}
{Mackey} D.,  et~al., 2019b, \mn@doi [\nat] {10.1038/s41586-019-1597-1}, \href {https://ui.adsabs.harvard.edu/abs/2019Natur.574...69M} {574, 69}

\bibitem[\protect\citeauthoryear{{Malhan} et~al.,}{{Malhan} et~al.}{2022}]{Malhan2022}
{Malhan} K.,  et~al., 2022, \mn@doi [\apj] {10.3847/1538-4357/ac4d2a}, \href {https://ui.adsabs.harvard.edu/abs/2022ApJ...926..107M} {926, 107}

\bibitem[\protect\citeauthoryear{{Massari}}{{Massari}}{2025}]{Massari2025}
{Massari} D.,  2025, \mn@doi [RNAAS] {10.3847/2515-5172/adc375}, \href {https://ui.adsabs.harvard.edu/abs/2025RNAAS...9...64M} {9, 64}

\bibitem[\protect\citeauthoryear{{Massari}, {Koppelman}  \& {Helmi}}{{Massari} et~al.}{2019}]{Massari2019}
{Massari} D.,  {Koppelman} H.~H.,   {Helmi} A.,  2019, \mn@doi [\aap] {10.1051/0004-6361/201936135}, \href {https://ui.adsabs.harvard.edu/abs/2019A&A...630L...4M} {630, L4}

\bibitem[\protect\citeauthoryear{{McConnachie} et~al.,}{{McConnachie} et~al.}{2018}]{McConnachie2018}
{McConnachie} A.~W.,  et~al., 2018, \mn@doi [\apj] {10.3847/1538-4357/aae8e7}, \href {https://ui.adsabs.harvard.edu/abs/2018ApJ...868...55M} {868, 55}

\bibitem[\protect\citeauthoryear{{Panithanpaisal}, {Sanderson}, {Wetzel}, {Cunningham}, {Bailin}  \& {Faucher-Gigu{\`e}re}}{{Panithanpaisal} et~al.}{2021}]{Panith2021}
{Panithanpaisal} N.,  {Sanderson} R.~E.,  {Wetzel} A.,  {Cunningham} E.~C.,  {Bailin} J.,   {Faucher-Gigu{\`e}re} C.-A.,  2021, \mn@doi [\apj] {10.3847/1538-4357/ac1109}, \href {https://ui.adsabs.harvard.edu/abs/2021ApJ...920...10P} {920, 10}

\bibitem[\protect\citeauthoryear{{Perina}, {Bellazzini}, {Buzzoni}, {Cacciari}, {Federici}, {Fusi Pecci}  \& {Galleti}}{{Perina} et~al.}{2012}]{Perina2012}
{Perina} S.,  {Bellazzini} M.,  {Buzzoni} A.,  {Cacciari} C.,  {Federici} L.,  {Fusi Pecci} F.,   {Galleti} S.,  2012, \mn@doi [\aap] {10.1051/0004-6361/201220037}, \href {https://ui.adsabs.harvard.edu/abs/2012A&A...546A..31P} {546, A31}

\bibitem[\protect\citeauthoryear{{Pietrinferni} et~al.,}{{Pietrinferni} et~al.}{2021}]{Pietrinferni2021}
{Pietrinferni} A.,  et~al., 2021, \mn@doi [\apj] {10.3847/1538-4357/abd4d5}, \href {https://ui.adsabs.harvard.edu/abs/2021ApJ...908..102P} {908, 102}

\bibitem[\protect\citeauthoryear{{Sakari} \& {Wallerstein}}{{Sakari} \& {Wallerstein}}{2022}]{Sakari2022}
{Sakari} C.~M.,  {Wallerstein} G.,  2022, \mn@doi [\mnras] {10.1093/mnras/stac752}, \href {https://ui.adsabs.harvard.edu/abs/2022MNRAS.512.4819S} {512, 4819}

\bibitem[\protect\citeauthoryear{{Sakari}, {Venn}, {Mackey}, {Shetrone}, {Dotter}, {Ferguson}  \& {Huxor}}{{Sakari} et~al.}{2015}]{Sakari2015}
{Sakari} C.~M.,  {Venn} K.~A.,  {Mackey} D.,  {Shetrone} M.~D.,  {Dotter} A.,  {Ferguson} A. M.~N.,   {Huxor} A.,  2015, \mn@doi [\mnras] {10.1093/mnras/stv020}, \href {https://ui.adsabs.harvard.edu/abs/2015MNRAS.448.1314S} {448, 1314}

\bibitem[\protect\citeauthoryear{{Sarajedini} et~al.,}{{Sarajedini} et~al.}{2007}]{sarajedini2007}
{Sarajedini} A.,  et~al., 2007, \mn@doi [\aj] {10.1086/511979}, \href {https://ui.adsabs.harvard.edu/abs/2007AJ....133.1658S} {133, 1658}

\bibitem[\protect\citeauthoryear{{Searle} \& {Zinn}}{{Searle} \& {Zinn}}{1978}]{SearleandZinn}
{Searle} L.,  {Zinn} R.,  1978, \mn@doi [\apj] {10.1086/156499}, \href {https://ui.adsabs.harvard.edu/abs/1978ApJ...225..357S} {225, 357}

\bibitem[\protect\citeauthoryear{{Stetson} et~al.,}{{Stetson} et~al.}{1999}]{Stetson1999}
{Stetson} P.~B.,  et~al., 1999, \mn@doi [\aj] {10.1086/300670}, \href {https://ui.adsabs.harvard.edu/abs/1999AJ....117..247S} {117, 247}

\bibitem[\protect\citeauthoryear{Tanvir et~al.,}{Tanvir et~al.}{2012}]{Tanvir2012}
Tanvir N.~R.,  et~al., 2012, \mn@doi [\mnras] {10.1111/j.1365-2966.2012.20590.x}, 422, 162

\bibitem[\protect\citeauthoryear{{Usher}, {Caldwell}  \& {Cabrera-Ziri}}{{Usher} et~al.}{2024}]{Usher2024}
{Usher} C.,  {Caldwell} N.,   {Cabrera-Ziri} I.,  2024, \mn@doi [\mnras] {10.1093/mnras/stae282}, \href {https://ui.adsabs.harvard.edu/abs/2024MNRAS.528.6010U} {528, 6010}

\bibitem[\protect\citeauthoryear{{VandenBerg}, {Brogaard}, {Leaman}  \& {Casagrande}}{{VandenBerg} et~al.}{2013}]{VDB2013}
{VandenBerg} D.~A.,  {Brogaard} K.,  {Leaman} R.,   {Casagrande} L.,  2013, \mn@doi [\apj] {10.1088/0004-637X/775/2/134}, \href {https://ui.adsabs.harvard.edu/abs/2013ApJ...775..134V} {775, 134}

\bibitem[\protect\citeauthoryear{{Veljanoski} et~al.,}{{Veljanoski} et~al.}{2014}]{Veljanoski14}
{Veljanoski} J.,  et~al., 2014, \mn@doi [\mnras] {10.1093/mnras/stu1055}, \href {https://ui.adsabs.harvard.edu/abs/2014MNRAS.442.2929V} {442, 2929}

\bibitem[\protect\citeauthoryear{{Villanova}, {Geisler}, {Carraro}, {Moni Bidin}  \& {Mu{\~n}oz}}{{Villanova} et~al.}{2013}]{Villanova2013}
{Villanova} S.,  {Geisler} D.,  {Carraro} G.,  {Moni Bidin} C.,   {Mu{\~n}oz} C.,  2013, \mn@doi [\apj] {10.1088/0004-637X/778/2/186}, \href {https://ui.adsabs.harvard.edu/abs/2013ApJ...778..186V} {778, 186}

\bibitem[\protect\citeauthoryear{{Weisz} et~al.,}{{Weisz} et~al.}{2016}]{Weisz2016}
{Weisz} D.~R.,  et~al., 2016, \mn@doi [\apj] {10.3847/0004-637X/822/1/32}, \href {https://ui.adsabs.harvard.edu/abs/2016ApJ...822...32W} {822, 32}

\bibitem[\protect\citeauthoryear{Zinn}{Zinn}{1993}]{Zinn93}
Zinn R.,  1993, The Globular Cluster-Galaxy Connection

\makeatother
\end{thebibliography}







\bsp	
\label{lastpage}
\end{document}